\documentstyle[latexsym,amssymb,aps,floats,eqsecnum,
preprint,amsfonts]{revtex}
\def\laq{\raise 0.4ex\hbox{$<$}\kern -0.8em\lower 0.62 ex\hbox{$\sim$}}
\def\gaq{\raise 0.4ex\hbox{$>$}\kern -0.7em\lower 0.62 ex\hbox{$\sim$}}
\input epsf
\begin{document}

\bibliographystyle{unsrt}

\title{Gauge field localization on Abelian vortices in six dimensions}

\author{Massimo Giovannini\footnote{Electronic address: 
massimo.giovannini@ipt.unil.ch} }

\address{{\it Institute of Theoretical Physics, 
University of Lausanne}}
\address{{\it BSP-1015 Dorigny, Lausanne, Switzerland}}

\maketitle
\begin{abstract}
The vector and tensor fluctuations of  vortices 
localizing gravity in the context of the six-dimensional Abelian 
Higgs model are  studied. These string-like solutions  
break spontaneously six-dimensional Poincar\'e invariance leading 
to a finite four-dimensional Planck mass and to a regular geometry 
both in the bulk and on the core of the vortex.
While the tensor modes of the metric 
are decoupled and exhibit a normalizable zero mode,
the vector fluctuations, present in the 
gauge sector of the theory, 
are naturally coupled to the graviphoton fields associated with 
the vector perturbations  
of the warped  geometry. Using the
invariance under infinitesimal diffeomorphisms,  it is found that 
 the zero modes of the  graviphoton fields are never localized. 
On the contrary,
the fluctuations of the Abelian gauge field itself admit a normalizable 
zero mode.  
\end{abstract}
\newpage
\renewcommand{\theequation}{1.\arabic{equation}}
\setcounter{equation}{0}
\section{Formulation of the problem}
Consider a $(4+2)$-dimensional space-time (consistent with four-dimensional 
Poincar\'e invariance) of the form \cite{m2} (see also \cite{ak,vis,seif})  
\begin{equation}
ds^2 = M^2(\rho)[ dt^2 - d\vec{x}^2] - d\rho^2 - L^2(\rho) d\theta^2,
\label{warp}
\end{equation}
where $\rho$ is the bulk radius,  $\theta$ is  the bulk angle. 
The specific form of the warp factors 
 $M(\rho)$ and $L(\rho)$ is determined by  consistency of 
the underlying theory of gravity with the generalized 
brane sources. 

In order to construct a  
theory incorporating gravitational and gauge interactions 
on warped geometries of the type of (\ref{warp}), 
 fields of various spin should be localized around 
the four-dimensional (Poincar\'e invariant) space-time.
Localized means  that the bulk fields exhibit  normalizable zero modes with 
respect to the  coordinates parametrizing the geometry in the 
transverse space. If the zero mode of a given fluctuation 
is not normalizable, then it will be decoupled from the 
four-dimensional physics.

A necessary condition for gravity localization on warped space-times
is a finite four-dimensional Planck mass. If the underlying 
gravity theory is the six-dimensional 
extension of the Einstein-Hilbert action, then, the 
four-dimensional Planck mass is, in the geometry of Eq. (\ref{warp}),
\begin{equation}
M_{P}^2 = 2 \pi M_{6}^4 \int_{0}^{\infty} M^2(\rho) L(\rho) d\rho,
\label{PM}
\end{equation}
where $M_{6}$ is the six-dimensional Planck mass.
The integral of Eq. (\ref{PM}) can be finite,  for large 
bulk radius,  if 
a (negative) cosmological constant is present in the bulk.
This situation is fully analogous to the 
five-dimensional case  where the effect of the 
bulk cosmological term is to give rise to an  ${\rm AdS}$
geometry for large  bulk radius  \cite{rs1,rs2}. Singularity 
free domain wall solutions in five dimensions can be also 
found \cite{gremm,tam} allowing to localize fields of spin lower than two 
\cite{tam}. 

Following the ideas put forward in the absence of gravitational 
interactions \cite{m2}, chiral fermionic degrees of freedom 
can be successfully localized in five-dimensions. 
The chiral fermionic zero mode is still present 
if the five-dimensional continuous space is 
replaced by a lattice \cite{kap}.
In six dimensions the situation is similar to the five-dimensional case
 but also different, since the structure of chiral zero modes may be more rich. 
The localization of fermionic degrees of freedom 
in six-dimensional (flat) space-time has been recently investigated
in \cite{tro1,tro2}. In \cite{tro1,tro2} an Abelian vortex 
plays the r\^ole of the scalar domain wall originally 
analyzed in \cite{m2}.

If  global topological defects 
are present together with a bulk cosmological 
constant, warped geometries leading to gravity localization
can be obtained \cite{def,def1,def2,def3}. A similar 
observation has been made in the context of  local defects \cite{gs}.

It has been recently shown \cite{mhm1,mhm2} that 
the Abelian-Higgs model  represents a well defined framework 
where local 
defects can lead to a six-dimensional geometry of the type of  (\ref{warp}).
For large bulk radius an  ${\rm AdS}_{6}$ space-time  
can be obtained.  
In this context the action is taken to be the six-dimensional 
generalization of the gravitating Abelian-Higgs model
\footnote{Conventions: Latin (uppercase) 
indices run over the six dimensional space-time. Greek indices run over 
the four (Poincar\'e invariant) dimensions.} 
\begin{equation}
S=\int
d^6x\sqrt{-G}\biggl[ - \frac{ R}{2 \kappa} - \Lambda+
\frac{1}{2}({\cal D}_{A}\varphi)^*{\cal D}^{A}\varphi-\frac{1}{4}
F_{AB}F^{AB}
-\frac{\lambda}{4}\left(\varphi^*\varphi-v^2\right)^2\biggr]~,
\label{a1}
\end{equation}
where ${\cal D}_{A}=\nabla_{A}-ieA_{A}$ is the gauge covariant derivative, 
while $\nabla_{A}$ is the generally covariant derivative. 
In Eqs. (\ref{a1}), $v$ is the vacuum expectation 
value of the Higgs field $\varphi$, $\lambda$ is the 
self-coupling constant and $e$ 
is the gauge coupling. Finally $\kappa = 8 \pi G_{6} \equiv 8 \pi /M_{6}^4$.

The action of Eq. (\ref{a1}) leads to equations of motion allowing 
static  solutions  
that depend only on the extra coordinates. Thus general
covariance along the four physical dimensions is unbroken. 
The corresponding 
background line element is  of the form (\ref{warp}) while 
the vortex ansatz for the gauge-Higgs system reads:
\begin{eqnarray} 
&& \varphi(\rho,\theta) =vf(\rho)e^{i\,n\,\theta},
\nonumber\\ 
&&A_{\theta}(\rho,\theta)  =\frac{1}{e}[\,n\,-\,P(\rho)] ~,
\label{NO}
\end{eqnarray}
where $n$ is the winding number. The local defect present in this 
theory  is the six-dimensional 
counterpart of the Abrikosov-type vortex arising in four dimensions \cite{no}. 
The radial and angular coordinates 
are replaced, in the present context, by the bulk radius and by the bulk
angle. 

There are different fields coming from the fluctuations of the  geometry 
which transform as divergence-less Poincar\'e vectors. These 
fields are garviphotons and will mix with the divergence-less fluctuations 
of the gauge sector leading to a non-trivial system determining the 
localization properties of the zero modes of the vector fluctuations 
of the model. This is the problem we ought to address. 

The vector fluctuations coming from the geometry change under 
infinitesimal coordinate transformations. This may lead to the 
unpleasant situation where the localization properties of a given field
change from one coordinate system to the other. Furthermore, it 
could also happen that in some cases spurious gauge modes appear in the game.
In order 
to avoid this problem we follow the approach already proposed and exploited  
\cite{mg1,mg2} in the 
analysis of five-dimensional domain-wall solutions \cite{tam,gremm}. 
The idea is to construct and use gauge-invariant
fluctuations which do not change under infinitesimal 
diffeomorphisms. The spirit of the analysis of \cite{mg1,mg2} 
was guided by the {\em Bardeen formalism} \cite{bard} whose useful 
features have been widely appreciated through the years in the 
context of (four-dimensional) cosmological models.  

In recent times various mechanisms have been put forward in order 
to localize vector fields in warped geometries 
(see \cite{rubr} for a nice review of the subject and 
\cite{tam} together with \cite{dv,od,rub,ti} for more detailed 
proposals). In \cite{tam} the localization of gauge fields 
is achieved through the coupling of the gauge kinetic term 
to a dilatonic field. In this example no background gauge field is present.
 In \cite{dv} the mechanism of localization is based on the assumption that the gauge theory is 
confining in the bulk but the confinement is absent on the brane. A 
realization of this scenario has been discussed in \cite{rub}. In \cite{ti}
a possible alternative to Higgs mechanism from higher dimensions has been 
discussed. As  byproduct of the analysis, 
the main ingredients for a successfull localization in five-dimensions 
have been listed. In all these models only the gauge field 
excitations have been considered. However, in a higher dimensional context 
vector modes certainly come from the metric excitations. Furthermore, 
the background gauge field is totally absent in these examples.

It should be clearly 
said that our considerations are not competitive with the level 
of generality of these mechanisms. The purpose of the present investigation is 
more specific. Given a class of vortex solutions localizing gravity 
in the well defined context of the Abelian-Higgs model, we ought to 
analyze systematically the vector and tensor excitations of the model. 
The six-dimensional Abelian-Higgs model with vortex solutions is 
interesting since the gauge field background is naturally 
present and it builds up, together with the Higgs field, the brane 
source. As a consequence, the structure of the vector zero modes is 
richer than in the case where the gauge field background is absent.

The plan of the paper is the following. In Section II the Abelian-Higgs
model will be discussed together with its vector fluctuations. In Section III
explicit vortex solutions leading to regular geometries localizing gravity  
will be introduced. In Section IV explicit evolution equations for the 
coupled system of graviphotons and gauge fluctuations 
will be derived and solved. The localization properties of the 
vector zero modes will be analyzed. Finally Section V contains the 
concluding remarks. Various technical results have been collected 
in the Appendix.

\renewcommand{\theequation}{2.\arabic{equation}}
\setcounter{equation}{0}
\section{Six-dimensional Abelian-Higgs models and its fluctuations}
From Eq.  (\ref{a1}) the related equations of motion  are:
 \begin{eqnarray}
&&G^{A B}\nabla_{A}\nabla_{B}\varphi
-e^2A_{A}A^{A} \varphi -ieA_{A}\partial^{A}\varphi-ie\nabla_{A}(A^{A}\varphi)
+\lambda (\varphi^*\varphi-v^2)\varphi=0,
\label{ph}\\
&&\nabla_{A} F^{AB}=-e^2A^{B}\varphi^*\varphi+\frac{ie}{2}
\left(\varphi\partial^{B}\varphi^*-\varphi^*\partial^{B}\varphi\right),
\label{A}\\
&& R_{AB}-\frac{1}{2}G_{AB}R = \kappa\left(T_{AB}+\Lambda G_{AB}\right),
\label{R}
\end{eqnarray}
where 
\begin{eqnarray}
T_{AB}&=&  \frac{1}{2}\left[({\cal D}_A\varphi)^*{\cal D}_{B}\varphi 
+({\cal D}_B\varphi)^*{\cal D}_A\varphi\right]
-\,F_{AC}{F_B}^{C} 
\nonumber\\
&-& G_{A B} \biggl[ \frac{1}{2}({\cal D}_{A}\varphi)^*
{\cal D}^{A}\varphi-\frac{1}{4}
F_{M N}F^{M N}-\frac{\lambda}{4}\left(\varphi^*\varphi-v^2\right)^2\biggr].
\label{enmom}
\end{eqnarray}

For the study of the fluctuations of the model it is often useful
to write Eq. (\ref{R}) in its contracted form where 
the scalar curvature is absent, namely:
\begin{equation}
R_{A B} = \kappa \tau_{A B}, 
\label{cont1}
\end{equation}
where 
\begin{equation}
\tau_{A B} = T_{A B} - \frac{T}{4} G_{A B} - \frac{\Lambda}{2} G_{A B}, ~~~~~T_{A}^{A} = T.
\label{cont2}
\end{equation}
Using Eq. (\ref{enmom}) into Eq. (\ref{cont2}), the explicit expression of $\tau_{AB}$ 
can be obtained: 
\begin{eqnarray}
\tau_{A B} &=& \biggl[ - \frac{\Lambda }{2}  + \frac{1}{8} F_{M N}F^{ M N} 
- \frac{\lambda }{8} ( \varphi^* \varphi - v^2)^2\biggr] G_{A B} 
\nonumber\\
&-& F_{A C}F^{~~~C}_{B} +
 \frac{1}{2} \left[({\cal D}_A\varphi)^*{\cal D}_{B}\varphi 
+({\cal D}_B\varphi)^*{\cal D}_A\varphi\right].
\label{cont3}
\end{eqnarray}

The fluctuation of the Abelian-Higgs model  
may arise both from the sources and from the metric. 
Notice that six-dimensional Poincar\'e invariance, as well as 
the local $U(1)$ symmetry, are naturally broken by the vortex 
ansatz. However, four-dimensional Poincar\'e invariance 
is still a good symmetry of the problem. It is 
therefore plausible to decompose 
the perturbations of the  metric  in terms of scalar vector and tensor fluctuations 
with respect to four-dimensional Poincar\'e transformations.
The generic metric fluctuation will then have 
scalar, vector and tensor modes, namely
\begin{equation}
\delta G_{AB}(x^{\mu}, w) = \delta G^{(S)}_{AB}(x^{\mu}, w) +\delta 
G^{(V)}_{AB}(x^{\mu}, w) +\delta G^{(T)}_{AB}(x^{\mu}, w),
\end{equation} 
which can be  parametrized as 
\begin{equation}
\delta G_{A B}=\left(\matrix{2M^2 H_{\mu\nu} 
& M {\cal G}_{\mu}  & L M {\cal B}_{\mu} \cr
 M {\cal G}_{\mu} & 2  \xi& L \pi\cr
L M {\cal B}_{\mu} & L \pi & 2 L^2 \phi&\cr}\right),
\label{lorf}
\end{equation}
where 
\begin{eqnarray}
&& H_{\mu\nu} = h_{\mu\nu} +\frac{1}{2} 
(\partial_{\mu} f_{\nu} +\partial_{\nu} f_{\mu}) 
+ \eta_{\mu \nu} \psi
+  \partial_{\mu}\partial_{\nu} E,
\nonumber\\
&& {\cal G}_{\mu} = D_{\mu} + \partial_{\mu} C,
\nonumber\\
&& {\cal B}_{\mu} = Q_{\mu} + \partial_{\mu} P,
\end{eqnarray}
with 
\begin{equation}
\partial_{\mu}h^{\mu}_{\nu} =0,~~~h_{\mu}^{\mu}=0,
\end{equation}
and with 
\begin{equation}
\partial_{\mu}f^{\mu} =0,~~~\partial_{\mu}D^{\mu} =0,~~~
\partial_{\mu}Q^{\mu} =0.
\end{equation}
Notice that $h_{\mu\nu}$ has five independent components, while  $Q_{\mu}$, $f_{\mu}$ and 
$D_{\mu}$ have, overall nine independent components. 
Finally $C$, $P$, $\psi$, $\phi$, $\xi$,  $E$ and $\pi$ correspond 
to seven scalar degrees of freedom.

The twenty one degrees of freedom of the perturbed six-dimensional metric 
change under infinitesimal coordinates transformations 
\begin{equation}
x^{A} \rightarrow \tilde{x}^{A} = x^{A} + \epsilon^{A}, 
\label{shift}
\end{equation}
as
\begin{equation}
\delta \tilde{G}_{A B} = \delta G_{AB} - \nabla_{A} \epsilon_{B} - \nabla_{B}
\epsilon_{A}, 
\label{liederiv}
\end{equation}
where
\begin{equation} 
\epsilon_{A} = ( M^2(\rho) \epsilon_{\mu}, -\epsilon_{\rho}, - L^2(\rho)\epsilon_{\theta}).
\end{equation}
In Eq. (\ref{liederiv}) the Lie 
the covariant derivatives are computed using the background metric
since the gauge transformations act around the fixed geometry 
defined by Eq. (\ref{warp}) compatible with the ansatz (\ref{NO}).
The infinitesimal shift $\epsilon_{\mu}$ along the four dimensional 
space-time can be decomposed, in its turn, as \footnote{Notice that 
since the bulk radius can go to infinity the decomposition (\ref{deca}) 
is well defined only if the scalar gauge function $\epsilon$ vanishes 
swiftly at infinity. In fact, from (\ref{deca}), 
$\Box \epsilon = \partial_{\mu} \epsilon^{\mu}$ which also 
implies that $\Box^{-1}$ exists if at infinity the gauge transformation is regular.
Thus, only regular gauge transformations will be discussed in the present context.}
\begin{equation}
\epsilon_{\mu} = \partial_{\mu}\epsilon + \zeta_{\mu}.
\label{deca}
\end{equation}
Hence, there will be two types of gauge transformations: the gauge 
transformations preserving the scalar nature of the fluctuations and the 
gauge transformations preserving the vector nature of the fluctuation.
The vector gauge transformations will involve pure vector gauge functions 
(i.e. $\zeta_{\mu}$) and will affect the three spin one fluctuations of the 
geometry (i.e. $f_{\mu}$, $D_{\mu}$ and $Q_{\mu}$). The scalar 
gauge transformations will involve pure scalar 
gauge functions (i.e. $\epsilon$, $\epsilon_{\rho}$ and $\epsilon_{\theta}$).

In the present investigation, only the vector and tensor modes 
of the geometry will be treated and the perturbed metric will 
then be \footnote{ The symbol $ \partial_{(\mu} f_{\nu)}$ denotes 
$(\partial_{\mu} f_{\nu} + \partial_{\nu} f_{\mu})/2$.}
\begin{equation}
\delta G_{A B}^{(T,V)}=\left(\matrix{2M^2 [ h_{\mu\nu} + \partial_{(\mu} f_{\nu)} ]
& M D_{\mu}  & L M Q_{\mu} \cr
 M D_{\mu} & 0& 0\cr
L M Q_{\mu} & 0& 0 &\cr}\right).
\label{lorf2}
\end{equation}
The transverse and traceless
tensors  are gauge-invariant, i.e. they do not change 
for infinitesimal gauge transformations
\begin{equation}
\tilde{h}_{\mu\nu} = h_{\mu\nu},
\label{hl}
\end{equation}
whereas the vector transform as 
\begin{eqnarray}
&& \tilde{f}_{\mu} = f_{\mu} - \zeta_{\mu},
\label{fl}\\
&&\tilde{D}_{\mu} = D_{\mu} - M\frac{\partial \zeta_{\mu}}{\partial \rho},
\label{zeta1}\\
&& \tilde{Q}_{\mu} = Q_{\mu} - \frac{M}{L} \frac{\partial\zeta_{\mu}}{\partial \theta}. 
\label{zeta2}
\end{eqnarray}
Since we have three vectors and one (vector) gauge function, two 
gauge-invariant vectors can be defined \cite{mg1,mg2},  corresponding
to six degrees of freedom
\begin{eqnarray}
&&\tilde{V}_{\mu} = \tilde{D}_{\mu} - M \frac{ \partial \tilde{f}_{\mu}}{\partial \rho},
\label{gia}\\
&& \tilde{Z}_{\mu} = \tilde{Q}_{\mu} - \frac{M}{L} \frac{\partial \tilde{f}_{\mu}}{\partial \theta}.
\label{gib}
\end{eqnarray}
Neither $V_{\mu}$ nor $Z_{\mu}$ change under 
infinitesimal gauge transformations. 

Around the fixed vortex background 
also the fluctuations of the source change for infinitesimal coordinate 
transformations:
\begin{equation}
\delta \tilde{A}_{A} = \delta A_{A} - A^{C} \nabla_{A} \epsilon_{C} 
- \epsilon^{B} \nabla_{B} A_{A}, 
\label{gshift}
\end{equation}
where $\delta A_{A}$ denotes the 
fluctuation of the gauge vector potential. The fluctuations 
$\delta A_{\rho}$ and $\delta A_{\theta}$ correspond to two scalar 
degrees of freedom. The fluctuation $\delta A_{\mu}$ can be simply decomposed, as 
\begin{equation}
\delta A_{\mu} = e {\cal A}_{\mu} + e \partial_{\mu} {\cal A}.
\label{decab}
\end{equation}
The gauge coupling has been introduced in the decomposition only for 
future convenience. The important point to be stressed is that since 
$\partial_{\mu} {\cal A}^{\mu} =0$, ${\cal A}$ transforms as a scalar and 
will not mix with the (divergence-less) vector modes of the geometry
defined previously. Since the 
only non-vanishing component of the vortex background (\ref{NO}) 
corresponds to $A_{\theta}$, then the pure vector 
fluctuation of the source, i.e. ${\cal A}_{\mu}$, will be automatically 
gauge-invariant i.e. , according to Eqs. (\ref{gshift}) and 
(\ref{decab}), 
\begin{equation}
{\tilde{\cal A}}_{\mu} = {\cal A}_{\mu}. 
\label{decac}
\end{equation}

The evolution equation for the tensor modes of the 
geometry is determined from the tensor component 
of Eq. (\ref{cont1}), namely \footnote{The symbol $\delta$ applied to a given 
tensor denotes 
the first order fluctuation of the corresponding quantity. } 
\begin{equation}
\delta R^{(T)}_{\mu\nu} = \kappa \delta \tau^{(T)}_{\mu\nu},
\label{I0}
\end{equation}
whereas the perturbed system of the vector  fluctuations is determined 
by the perturbed Einstein's equations carrying vector indices 
\begin{eqnarray}
&& \delta R^{(V)}_{\mu \nu} = \kappa \delta \tau^{(V)}_{\mu \nu},
\label{I}\\
&& \delta R^{(V)}_{\mu \rho} = \kappa \delta \tau^{(V)}_{\mu \rho},
\label{II}\\
&&  \delta R^{(V)}_{\mu \theta} = \kappa \delta \tau^{(V)}_{\mu \theta},
\label{III}
\end{eqnarray}
supplemented  by the the perturbed vector component of the gauge field equation (\ref{A})
\begin{eqnarray}
&& \delta G^{ A C} \biggl[ \partial_{C} F_{A \mu} - 
\overline{\Gamma}_{ A C}^{D} F_{D\mu} -
\overline{\Gamma}_{\mu C}^{D} F_{A D} \biggr] + e^2 \delta A_{\mu} \varphi^* \varphi
\nonumber\\
&& G^{ AC}[ \partial_{C} \delta F_{A \mu} - \delta \Gamma_{ A C}^{D} 
F_{D \mu} - 
\Gamma_{A C}^{D} \delta F_{D \mu} - \delta \Gamma_{ \mu C}^{D} F_{ A D} 
- \Gamma_{B C}^{D} \delta F_{AD} \biggr] =0,
\label{Aper}
\end{eqnarray}
In the previous equations $\delta$ denotes the first order fluctuation 
of the corresponding quantity. In Eqs. (\ref{I})--(\ref{III}) 
$\delta \tau_{A B}$ represents the fluctuations of Eq. (\ref{cont3}) and 
\begin{equation}
\delta R_{AB} = \partial_{C} \delta \Gamma_{A B}^{C} - \partial_{B} \delta 
\Gamma_{A C}^{C} + \overline{\Gamma}_{A B}^{C} \delta \Gamma_{ C D}^{D} 
+  \delta\Gamma_{A B}^{C} \overline{\Gamma}_{ C D}^{D}- 
\delta\Gamma_{B C}^{D} \overline{\Gamma}_{A D}^{C} - 
\overline{\Gamma}_{B C}^{D} \delta\Gamma_{A D}^{C}. 
\label{ricci}
\end{equation}
In Eq. (\ref{ricci}), $\overline{\Gamma}_{ A B}^{C}$ are the background 
values of the Christoffel connections. In Appendix A all the 
explicit values of the fluctuations are reported for the case under study.

\renewcommand{\theequation}{3.\arabic{equation}}
\setcounter{equation}{0}
\section{Vortex Solutions in warped spaces}
In this Section the main properties of the vortex solutions will be 
outlined. Explicit solutions will also be presented.
Inserting Eqs.(\ref{warp}) and  (\ref{NO}) into Eqs. (\ref{ph})--(\ref{R})
and using the following rescalings for the parameters of the model
\footnote{Notice that 
the Higgs boson and vector masses are, in our definitions, 
$m_{H} = \sqrt{2 \lambda} ~ v$ and $ m_{V} = e v$.} 
\begin{equation}
\nu = \kappa v^2, ~~~~~ \alpha = \frac{e^2}{\lambda} ,~~~
\mu = \frac{\kappa \Lambda}{\lambda v^2}.
\label{def}
\end{equation}
we get the background equations of motion in their explicit form:
\begin{eqnarray} 
&& f'' + ( 4 H + F ) f' +(1 - f^2) f -
\frac{P^2}{{\cal L}^2}f=0,
\label{f1}\\ 
&& P''  + ( 4 H - F) P' 
  -\alpha f^2 P=0,
\label{p1}\\ 
&& F' + 3 H' 
+ F^2 + 6 H^2 + 3 H F  = - \mu - \nu \biggl[\frac{ {f'}^2}{2} 
+ \frac{1}{4} ( f^2 -1 )^2 
+ \frac{{P'}^2}{2 \alpha {\cal L}^2 }  
+ \frac{f^2 P^2 }{2 {\cal L}^2}\biggr] ,
\label{m1}\\
&& 4  H'  + 10 H^2 = - \mu - \nu\biggl[
 \frac{{f'}^2}{2} 
+ \frac{1}{4} ( f^2 -1 )^2 
- \frac{{P'}^2}{2 \alpha {\cal L}^2 }  
- \frac{f^2 P^2 }{2 {\cal L}^2}\biggr] ,
\label{m2}\\
&& 4 H F + 6 H^2 = -\mu - \nu \biggl[ -\frac{{f'}^2}{2} 
+ \frac{1}{4} ( f^2 -1 )^2 
- \frac{{P'}^2}{2 \alpha {\cal L}^2 }  
+ \frac{f^2 P^2 }{2 {\cal L}^2}  \biggr]. 
\label{l1}
\end{eqnarray}
In Eqs. (\ref{f1})--(\ref{l1}) 
the prime denotes the derivation with respect to 
the rescaled variable 
\begin{equation}
 x = \,m_{H}\,\rho/\sqrt{2} \equiv \sqrt{\lambda} v 
\rho,
\end{equation}
and the  function $L(\rho)$ appearing in the line element of Eq. 
(\ref{warp}) has also been rescaled, namely
\begin{equation}
{\cal L}(x) = \sqrt{\lambda} v L(\rho).
\end{equation}
In Eqs. (\ref{f1})--(\ref{l1}), $H$ and $F$ denote the derivatives 
(with respect to $x$)
of the logarithms of the warp factors:
\begin{equation}
H(x) = \frac{d \ln{M(x)}}{d x},\,\,\,
F(x) = \frac{d \ln{{\cal L}(x)}}{d x}.
\end{equation}

\begin{figure}
\centerline{\epsfxsize = 10 cm  \epsffile{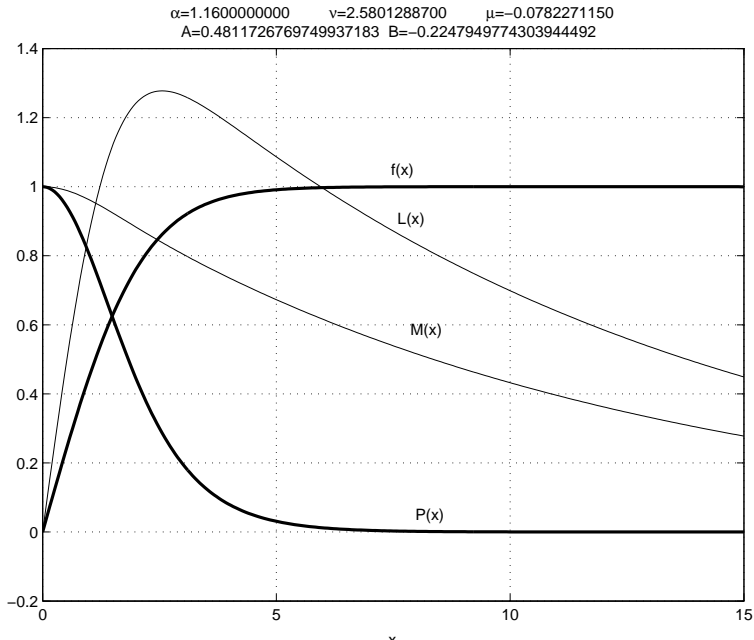}} 
\caption[a]{The vortex solution. The parameters chosen for this example lie 
on the surface defining the parameter space of the model and illustrated in Fig. \ref{F4}.}
\label{F1}
\end{figure}

\begin{figure}
\centerline{\epsfxsize = 10 cm  \epsffile{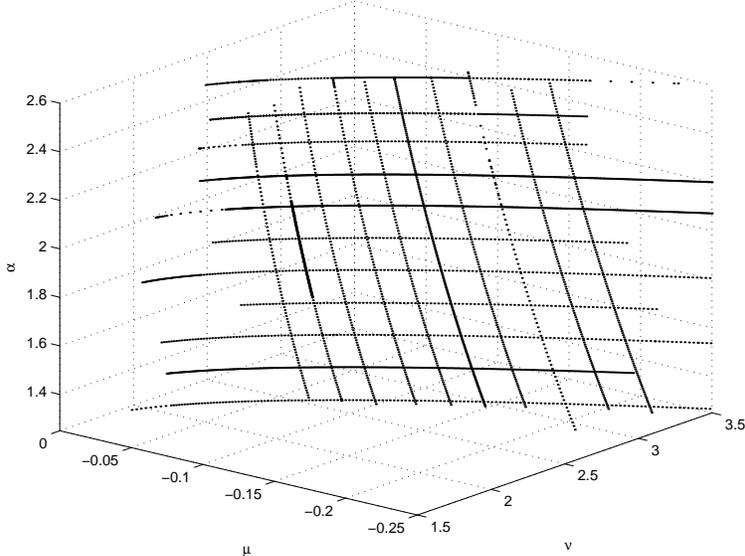}} 
\caption[a]{The parameter space of the vortex solution is illustrated in terms of 
the dimension-less parameters reported in Eq. (\ref{def}).}
\label{F4}
\end{figure}

The solution reported in Fig. \ref{F1}
 is representative of a class of 
solutions whose parameter space is illustrated in Fig. \ref{F4}
in terms of the dimension-less parameters of Eq. (\ref{def}). 
From Fig. \ref{F1}, recalling the vortex ansatz of Eq. (\ref{NO}), 
  the scalar field reaches, for large $x$, its vacuum expectation  
value, namely
$|\phi(\rho)| \rightarrow v$  for $\rho \rightarrow \infty$. In the
same limit, the gauge field goes to zero. Close to
the core of the string  both fields are  regular. These
properties of the solutions 
can be translated in terms of our  rescaled variables as
\begin{eqnarray} 
f(0)=0,&\qquad& \lim_{x\rightarrow \infty} f(x)=1,
\nonumber\\ 
P(0)=n,&\qquad& \lim_{x\rightarrow \infty} P(x)=0. 
\label{boundary}
\end{eqnarray}
Notice that the solutions of Fig. \ref{F1} and \ref{F4} correspond to the 
case of lowest winding\footnote{The vortex solutions presented in this Section can be also 
generalized to the case of higher winding, i.e. $n \geq 2 $ \cite{mhm1}.}, 
i.e. $n=1$ in Eq. (\ref{NO}).
The requirement of regular geometry in the core of the string reads
\begin{equation}
\left.\frac{ d M}{d x}\right|_{0} 
= 0~,~~{\cal L}(0) = 0~,~~ \left.\frac{d{\cal L}}{d x}
\right|_{0} = 1~,
\label{boundary2}
\end{equation}
and $M(0) =1$. 
More specifically, at large distances from the core the 
behaviour of the geometry is  ${\rm AdS}_{6}$ space-time
characterized  by exponentially 
decreasing warp factors
\begin{equation}
M(x) \sim e^{ - c x },~~~~{\cal L}(x) \sim e^{- c x}, 
\label{ads}
\end{equation}
where $ c = \sqrt{- \mu/10}$. This behaviour can 
be understood since the defects corresponding to the solution
of Fig. \ref{F1} are {\em local} and their related
energy-momentum tensor goes to zero at large 
distances where the geometry is determined 
only by the value of the bulk cosmological constant.
The form of the solutions in the vicinity of the core of the vortex
can be  studied by  expressing the metric functions together with
the  scalar and gauge fields  as a power series in $x$, the
dimensionless bulk radius.  The power series will then be inserted
into Eqs.  (\ref{f1})--(\ref{l1}). Requiring that the series obeys,
for $x\rightarrow 0$,  the boundary conditions of Eqs.
(\ref{boundary})  the form of the solutions can be determined as a
function of the  parameters of the model:
\begin{eqnarray}
f(x)&\simeq&Ax+\left(\frac{2\mu}{3}+\frac{\nu}{6}+\frac{4B^2 
\nu}{3\alpha}+\frac{2A^2 \nu}{3}-1+2B\right)\frac{A}{8}x^3,\\
P(x)&\simeq&1+Bx^2,\\
M(x)&\simeq&1+\left(-\frac{\mu}{8}-\frac{\nu}{32}+
\frac{\nu B^2}{4\alpha}\right)x^2,\\
{\cal L}(x)&\simeq&x+\left[\frac{\mu}{12}+\nu(\frac{1}{48}
-\frac{5B^2}{6\alpha}-\frac{A^2}{6})\right]x^3.
\label{x=0}
\end{eqnarray}
In Eq. (\ref{x=0}) $A$ and $B$ are two arbitrary constants which cannot 
be determined by the local analysis of the equations of motion. These
constants are to be found by boundary conditions for $f(x)$ and
$P(x)$ at infinity. 

\begin{figure}
\centerline{\epsfxsize = 10 cm  \epsffile{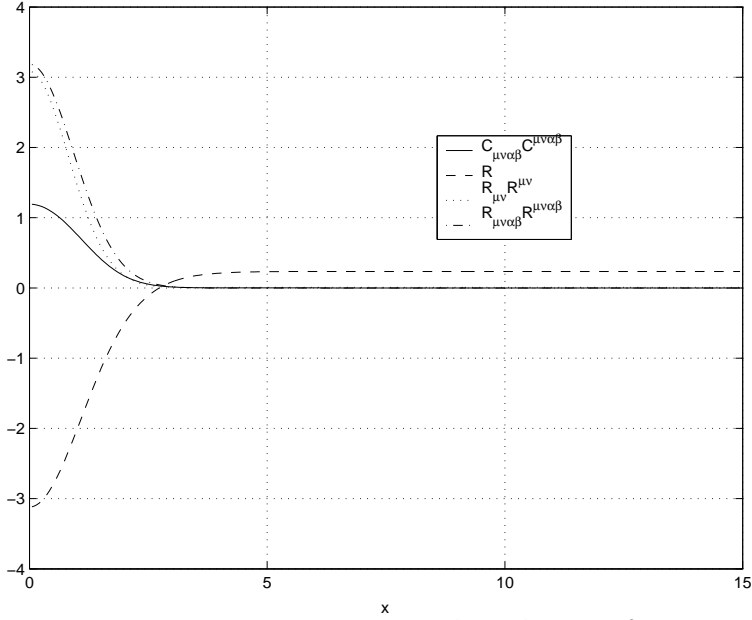}} 
\caption[a]{The curvature invariants pertaining to the 
solution of Fig. \ref{F1}. All the curvature invariants are regular 
for the class of vortex solutions examined in the 
present analysis.}
\label{F2}
\end{figure}

By studying the relations among the string tensions 
it is possible tro determine the value of $B$ as discussed in detail 
in \cite{mhm1}. 
For all the solutions of the family defined by the fine-tuning surface of 
Fig. \ref{F4} we have that 
\begin{equation}
-\frac{\nu}{\alpha {\cal L}}\left.\frac{d P}{ d x }\right|_0=1\,.
\label{mus2}
\end{equation}
Since $dP/dx$ is related to the magnetic field on the vortex, 
Eq. (\ref{mus2}) tells that in order to have ${\rm AdS}_{6}$ (at infinity)
and a local vortex (around the origin) a specific relation among the 
dimension-less couplings 
and the value of the magnetic field on the vortex must hold.
In fact, according to Eq. (\ref{x=0}), for $x \rightarrow 0$, 
$ P\sim 1 + B x^2$. Using 
Eq. (\ref{mus2}), the expression for $B$ can be exactly computed
\begin{equation}
B= - \frac{\alpha}{2\nu}\,.
\label{B}
\end{equation}
The relation (\ref{B})  among the parameters is satisfied by all the solutions 
of the type of Fig. \ref{F1}. This aspect can be appreciated 
by looking at the specific numerical values of the different 
parameters reported on top of the plots.
The solutions belonging to the family represented by Fig. \ref{F1} 
are regular everywhere, not only in the origin or at infinity. 
This feature of the solution is illustrated by the  
behaviour of the curvature invariants which are reported in Fig. \ref{F2}.
While the vortex solutions have been analytically obtained, 
the asymptotic behaviour of the gauge and Higgs field can be 
analytically understood in simple terms.

If we expand the gauge and Higgs fields around their 
boundaries
\begin{eqnarray}
&& P(x) = \overline{P} + \delta P(x),~~~ \overline{P}\sim 0,
\nonumber\\
&& f(x) = \overline{f} - \delta f(x),~~~ \overline{f}=1,
\end{eqnarray}
 and if we take into account that, in the same limit, the warp factors
 decrease 
exponentially, then we get from Eqs. (\ref{f1})--(\ref{l1}) 
\begin{eqnarray}
&& \delta P(x) \sim e^{ \sigma_{1} x},\,\,\sigma_1 = 
\frac{3 c}{2} \bigl[ 1 \pm \sqrt{1+  \frac{4 \alpha}{9 c^2}}\bigr],
\label{dP}
\nonumber\\
&& \delta f(x) \sim e^{\sigma_2 x},\,\,\,\, \sigma_2 = 
\frac{5 c}{2} \bigl[ 1 \pm \sqrt{ 1 + \frac{8}{25 c^2}}\bigr].
\label{df}
\end{eqnarray}
If  $4\alpha \gg 9 c^2$ (the limit of small bulk cosmological
constant) the solution  is compatible with the gauge field decreasing
asymptotically as  $\delta P \sim e^{-\sqrt{\alpha}x}$.
If  $ 25 c^2 < 8$ the the perturbed  solution goes as  $\delta f \sim
e^{- \sqrt{2} x}$. 

There is an interesting relations which may be derived by direct 
integration of the equations of motions.
Consider  the difference between the $(0,0)$ and $(\theta,\theta)$
components of the  Einstein equations, i.e. (\ref{m1}) and (\ref{m2}):
\begin{equation}
\frac{ d (H - F)}{ d x} 
+ ( F + 4 H) (F - H) = - \nu( \tau_0 - \tau_{\theta} ) 
\end{equation}
Multiplying both sides of this equation by $M^4 {\cal L}$ and integrating 
from $0$ to $x$ we get 
\begin{equation}
( H - F)  = \frac{\nu}{\alpha} 
\frac{ P }{{\cal L}^2}\frac{ d P }{d x} - 
\biggl( 1 + \left.
 \frac{\nu}{\alpha {\cal L}} \frac{d P}{d x }
\right|_0\biggr) \frac{1}{M^4 {\cal L}}.
\label{tuneq}
\end{equation}
If the tuning among the string tensions is enforced, according to 
Eqs. (\ref{mus2})
and (\ref{B}) the boundary term 
in the core disappears and the resulting equation will be 
\begin{equation}
F = H - \frac{\nu}{\alpha} \frac{P~}{{\cal L}^2} \frac{ d P}{d x}.
\label{fminh}
\end{equation}
This relation holds for all the family of solutions 
describing vortex-type solutions with 
${\rm AdS}_{6}$ behaviour at infinity and it will turn out to be 
important for the analysis of the localization properties of 
the vector fluctuations of the model.

\renewcommand{\theequation}{4.\arabic{equation}}
\setcounter{equation}{0}
\section{Tensor and vector fluctuations of the gravitating vortex}
\subsection{Tensor fluctuations}
The evolution equation for the spin two fluctuations 
of the geometry are decoupled from the very beginning and 
they are easily obtained from the tensor component
of the perturbed Einstein equations, namely from Eq. (\ref{I0}).
The details are reported in Appendix A 
and the final result is \footnote{Since the bulk space-time has two transverse coordinates (i.e. 
$x =\sqrt{\lambda} \rho$) and $\theta$, the derivative with respect to $\theta$ will 
be denoted by an overdot, whereas the derivative with respect to $x$ will 
be denoted, as usual,  by a prime.}
\begin{equation}
h_{\mu\nu}'' + \frac{1}{{\cal L}^2} \ddot{h}_{\mu\nu} + 
(4 H + F) h_{\mu\nu}' - \frac{2}{m_{H}^2 M^2} \Box h_{\mu\nu} =0.
\label{tens}
\end{equation}
For the lowest mass eigenstate (i.e. $\Box h_{\mu\nu} =0$ and $\ddot{h}_{\mu\nu} =0$), 
Eq. (\ref{tens}) admit a solution with constant amplitude. 
Denoting with $h$ each polarization, the 
 constant zero mode will be $h = K$ where $K$ is a constant. 

In order to discuss 
the localization properties of the zero mode, the canonically 
normalized kinetic term (coming from the action 
perturbed to second order) should be derived. Going to the action 
of the tensor fluctuation the canonical variable 
\begin{equation}
q = M \sqrt{L} h,
\label{cantens}
\end{equation}
can be read-off. In terms of $q$ the kinetic term of the 
tensor fluctuations is canonically normalized and the 
corresponding normalization integral is then 
\begin{equation}
K^2 \int_{0}^{\infty} M^2(x) {\cal L}(x) d x. 
\label{normt}
\end{equation}
The requirement that the integral (\ref{normt}) 
is finite corresponds to the requirement that 
the four-dimensional Planck mass is finite. In fact, Eq. (\ref{PM}) can be also 
written as  
\begin{equation}
M_{P}^2 = \frac{4 \pi}{m_{H}^2} 
M^4_{6} \int_{0}^{\infty} dx M^2(x) {\cal L}(x).
\label{PM2}
\end{equation}
Therefore, if the four-dimensional Planck mass is finite 
also the tensor zero mode is normalizable. Furthermore, 
using the asymptotics of the background solutions reported in Eqs. (\ref{x=0}) and 
(\ref{ads}) the following behaviour of the canonically normalized 
zero mode can be obtained:
\begin{eqnarray}
&& q(x) \sim  e^{ - \frac{3}{2} c x} , ~~~~~x \to \infty
\nonumber\\
&& q(x) \sim  \sqrt{x}, ~~~~~~~~~~x \to 0.
\label{tzm}
\end{eqnarray}
As a consequence, the normalization integral (\ref{normt}) as well as 
(\ref{PM2}) will always give finite results.
 These findings are in full agreement with 
the ones reported in \cite{gs} with the difference that, in the present case, 
we are dealing with an explicit model of vortex based on the Abelian-Higgs model. 
The reason why tensor fluctuations can be analyzed even without an
explicit model, is that they decouple mfrom the sources. The same 
statement cannot be made for the vector modes of the geometry whose 
coupling to the sources is essential,  as it will be now 
discussed.

\subsection{Vector fluctuations}

In terms of the gauge-invariant quantities defined in Section II
the evolution equations for the vector modes of the 
system can be derived from Eqs. (\ref{I})--(\ref{III}) and from Eq. 
(\ref{Aper}). While the detailed derivation is reported in Appendix 
A, the final result of the straightforward but lengthy 
algebra is the following
\begin{eqnarray}
&& V_{\mu}' + ( 3 H + F) V_{\mu}  + \frac{\dot{Z}_{\mu}}{{\cal L}} =0,
\label{v1d}\\
&& \frac{\ddot{V}_{\mu}}{{\cal L}^2} - \frac{2}{m_{H}^2 ~M^2}\Box V_{\mu} - 
\frac{\dot{Z}_{\mu}'}{{\cal L}} + ( H - F) \frac{\dot{Z}_{\mu}}{{\cal L}} = 
2 \frac{L}{M} \frac{\nu}{\alpha} \frac{P'}{{\cal L}^2} 
\frac{\dot{{\cal A}}_{\mu}}{{\cal L}},
\label{v2d}\\
&& Z_{\mu}'' + ( 4 H + F) Z_{\mu}' + ( F' - H' + 5 H F - 5 H^2) Z_{\mu} 
- \frac{ 2}{ m_{H}^2 ~M^2} \Box Z_{\mu}
\nonumber\\
&& - \frac{1}{{\cal L}} \biggl[ \dot{V}_{\mu}' + ( 5 H 
- F) \dot{V}_{\mu}\biggr] + 2 \frac{L}{M} \biggl( \frac{\nu}{\alpha} 
\frac{P'}{{\cal L}^2} {\cal A}_{\mu}' + \frac{\nu P f^2}{{\cal L}^2} 
{\cal A}_{\mu} \biggr) =0,
\label{v3d}\\
&& {\cal A}_{\mu}'' 
+ \frac{\ddot{{\cal A}}_{\mu}}{{\cal L}^2} - \frac{2}{m_{H}^2~M^2} \Box
{\cal A}_{\mu} + ( 2 H + F) {\cal A}_{\mu}'
\nonumber\\
&& - P' \frac{M}{L} \biggl[ Z_{\mu}' - (H- F) Z_{\mu} - \frac{\dot{V}_{\mu}}{{\cal L}}
\biggr] - \alpha {\cal A}_{\mu} f^2 =0.
\label{v4d}
\end{eqnarray}
This system defines the evolution of the three gauge-invariant vector 
fluctuations appearing in the gravitating Abelian Higgs model, namely $V_{\mu}$, 
$Z_{\mu}$ and ${\cal A}_{\mu}$. These three vectors are divergence-less 
and have been defined in Eqs. (\ref{zeta1})--(\ref{zeta2}) and in eq. (\ref{decac}).
While Eq. (\ref{v1d}) is a constraint, the other equations are all dynamical.

In order to simplify the above system it is useful to introduce the following 
combination of the derivatives of the two graviphoton fields namely:
Define now the following variable
\begin{equation}
u_{\mu} = \varepsilon \frac{\dot{V}_{\mu}}{{\cal L}} - ( \varepsilon Z_{\mu})',
\label{u}
\end{equation}
where $\varepsilon$ is a background function satisfying
\begin{equation}
\varepsilon = \frac{L}{M},~~~~~~~\frac{\varepsilon'}{\varepsilon} = F - H.  
\end{equation}
Using Eq. (\ref{u}), Eqs. (\ref{v1d})--(\ref{v4d}) 
can then  be written in a more compact form, namely: 
\begin{eqnarray}
&& V_{\mu}' + ( 4 H + \frac{\varepsilon'}{\varepsilon}) V_{\mu} + 
\frac{\dot{Z}_{\mu}}{{\cal L}}=0,
\label{v1e}\\
&& \frac{\dot{u}_{\mu}}{{\cal L}} - \frac{2}{m_{H}^2} 
\frac{\varepsilon}{M^2} \Box V_{\mu} = 2 \varepsilon^2 \frac{\nu}{\alpha} 
P' \frac{\dot{{\cal A}}_{\mu}}{{\cal L}},
\label{v2e}\\
&& u_{\mu}' + ( 5 H - \frac{\varepsilon'}{\varepsilon}) u_{\mu} + 
2 \frac{\varepsilon}{m_{H}^2 M^2} \Box Z_{\mu} = 2 \varepsilon^2 \biggl[ 
\frac{\nu}{\alpha} \frac{P'}{{\cal L}^2} {\cal A}_{\mu}' + \frac{\nu P f^2}{{\cal L}^2} 
{\cal A}_{\mu}\biggr],
\label{v3e}\\
&& 
{\cal A}_{\mu}'' + \frac{\ddot{{\cal A}}_{\mu}}{{\cal L}^2} 
- \frac{2}{m_{H}^2 ~M^2} 
\Box {\cal A}_{\mu} + ( 2 H + F) {\cal A}_{\mu}' 
+ \frac{P'}{\varepsilon^2}u_{\mu} - 
\alpha f^2 {\cal A}_{\mu}=0.
\label{v4e}
\end{eqnarray}
In order to determine the zero modes of the system consider first the case 
where the mass of $V_{\mu}$ vanishes, i.e. $\Box V_{\mu} =0$. In this case from Eq. 
(\ref{v2e}) the following relation holds:
\begin{equation}
u_{\mu} = 2 \varepsilon^2 \frac{\nu}{\alpha}  \frac{P'}{{\cal L}^2} 
{\cal A}_{\mu}. 
\label{umeq}
\end{equation}
Inserting Eq. (\ref{umeq}) into Eq. (\ref{v4e}) 
the equation for the gauge field fluctuation can be 
simplified with the result  that 
\begin{equation}
{\cal A}_{\mu}'' + \frac{\ddot{{\cal A}}_{\mu}}{{\cal L}^2} 
- \frac{2}{m_{H}^2 ~M^2} 
\Box {\cal A}_{\mu} + ( 2 H + F) {\cal A}_{\mu}' 
+ \biggl[ 2 \frac{\nu}{\alpha} \frac{{P'}^2}{{\cal L}^2}  - 
\alpha f^2 \biggr]{\cal A}_{\mu}=0.
\label{int1}
\end{equation}
Already from this expression we can see  that the zero mode 
of ${\cal A}_{\mu}$ will not be constant but it will be a function 
of the bulk radius. This property is in contrast with 
the results obtained in the case of the tensor zero mode.
In order to solve explicitely for the zero mode 
Eq. (\ref{int1}) can be further simplified by using the equations of motion of the 
background. In particular, from Eq. (\ref{f1}) we have 
that 
\begin{equation}
\alpha f^2 = \frac{P''}{P} + ( 4 H - F) \frac{P'}{P}, 
\label{int2}
\end{equation}
whereas, using Eq. (\ref{fminh}) we also have that 
\begin{equation}
2 \frac{\nu}{\alpha} \frac{{P'}^2}{{\cal L}^2} = ( H -F) \frac{P'}{P}.
\label{int3}
\end{equation}
Inserting Eqs. (\ref{int2}) and (\ref{int3}) into the last term of Eq. 
(\ref{int1}) the following equation can be obtained, namely
 \begin{equation}
{\cal A}_{\mu}'' + \frac{\ddot{{\cal A}}_{\mu}}{{\cal L}^2} 
- \frac{2}{m_{H}^2 ~M^2} 
\Box {\cal A}_{\mu} + ( 2 H + F) {\cal A}_{\mu}' 
- \biggl[\frac{P''}{P} + ( 2 H + F) \frac{P'}{P}  \biggr]{\cal A}_{\mu}=0.
\label{int4}
\end{equation}
Finally, defining the appropriately rescaled variable,
\begin{equation}
b_{\mu} = M \sqrt{L} {\cal A}_{\mu}
\end{equation}
the first derivative with respect to the bulk radius can be eliminated from Eq. (\ref{int4})
with the result that 
\begin{equation}
b_{\mu}'' + \frac{\ddot{b}_{\mu}}{{\cal L}^2} - \frac{2}{m_{H}^2 M^2 } \Box b_{\mu} 
- \frac{ (M \sqrt{L} P)''}{M \sqrt{L} P} b_{\mu} =0,
\label{zerog}
\end{equation}
From Eq. (\ref{zerog}) the corresponding 
zero mode can be easily obtained.
 Expanding the excitation in Fourier series with respect to 
$\theta$ and in Fourier integral with respect to the four-momentum 
\begin{equation}
b_{\mu}(x^{\mu},\theta, w) = \sum_{\ell= -\infty}^{+\infty} \int e^{i p x} d^{4} p~ b_{\mu}(\ell, p, w).
\end{equation}
Hence, in Eq. (\ref{zerog}) the D'Alembertian is replaced by $-m^2$ and the term containing the double derivative 
with respect to $\theta$ is replaced by $-\ell^2$. The lowest mass and angular momentum eigenstates
will then obey the following equation for the zero mode   
\begin{equation}
b_{\mu}''  
- \frac{ (M \sqrt{L} P)''}{M \sqrt{L} P} b_{\mu} =0.
\end{equation}
whose solution, in terms of ${\cal A}_{\mu}$ can be written as 
\begin{equation}
 {\cal A}_{\mu}(x) = k_{1,\mu} P(x),
\label{zmp}
\end{equation}
where $k_{1,\mu}$ is an integration constant 
which should be determined by normalizing the canonical 
zero mode associated with gauge field fluctuations. As 
anticipated this zero mode is a specific function 
of the bulk radius and not a constant.
More specifically, according to Eqs. (\ref{x=0})--(\ref{B}) and (\ref{dP}),
\begin{eqnarray} 
&&{\cal A}_{\mu}(x) \simeq 1 - \frac{\alpha}{2\nu} x^2,~~~~  x\to 0,
\nonumber\\ 
&&{\cal A}_{\mu}(x) \simeq e^{- \sqrt{\alpha}x},~~~~x \to\infty.
\end{eqnarray}
Hence, the zero mode of Eq. (\ref{zmp}) satisfies the correct boundary 
conditions since 
\begin{equation}
{\cal A}_{\mu}'(0) = {\cal A}_{\mu}'(\infty) =0,
\end{equation}
so that the differential operator of Eq. (\ref{int4}) is self-adjoint.

Inserting Eq. (\ref{umeq}) into Eq. (\ref{v3e})( or (\ref{v3d}), 
the obtained equation implies that 
\begin{equation}
\Box Z_{\mu} =0,
\end{equation}
so that also the second graviphoton field should have zero mass. 
In order to determine the explicit expressions of the zero modes related to $V_{\mu}$ 
and $Z_{\mu}$ we should consider radial excitations. Thus, 
inserting Eq. (\ref{zmp}) into Eq. (\ref{umeq}) and recalling 
Eq. (\ref{u}) the zero mode for $Z_{\mu}$ can be obtained 
\begin{equation}
Z_{\mu}(x) =   k_{1,\mu} \frac{L(x)}{M(x)} + c_{2,\mu} \frac{M(x)}{L(x)},
\label{0q}
\end{equation}
where $c_{2,\mu}$ is a further integration constant.
From Eq. (\ref{v1e}) the zero mode of $V_{\mu}$ turns out to be
\begin{equation}
 V_{\mu}(x) = \frac{c_{1,\mu}}{M^3(x) L(x)},
\label{0d}
\end{equation}
where $c_{1,\mu}$ is the integration constant.

By  now perturbing the action to second order the correct canonical 
normalization of the fields can be deduced.  The canonical fields 
related to ${\cal A}_{\mu}$, $V_{\mu}$ and $Z_{\mu}$ are 
\begin{eqnarray}
&&\overline{{\cal A}}_{\mu} = \sqrt{L} {\cal A}_{\mu},
\nonumber\\
&&\overline{V}_{\mu} = M\sqrt{L} V_{\mu},
\nonumber\\
&&\overline{Z}_{\mu} = M \sqrt{L} Z_{\mu}. 
\label{cannm}
\end{eqnarray}
In order to get to Eqs. (\ref{cannm}), 
it is better to perturb to second order 
the Einstein-Hilbert action  directly in the form 
\cite{landau}
\begin{equation}
G^{A B}\biggl( \Gamma_{A C}^{D} \Gamma_{B D}^{C}
- \Gamma_{A B}^{C} \Gamma_{C D}^{D}\biggr) 
\end{equation}
where the total derivatives are absent.
Using the results of Section III, and, in particular, 
Eqs. (\ref{x=0}) and (\ref{dP}) the canonically 
normalized gauge zero mode behaves, asymptotically, as 
\begin{eqnarray}
&& \overline{{\cal A}}_{\mu}(x) \sim   e^{ - ( \frac{c}{2} + \sqrt{\alpha})x}, ~~~~~~x \to \infty,
\nonumber\\
&& \overline{{\cal A}}_{\mu}(x) \sim  \sqrt{x}, ~~~~~x\to 0
\label{zmgas}
\end{eqnarray}

The canonically normalized graviphoton fields will 
behave, asymptotically, as 
\begin{eqnarray}
&& \overline{V}_{\mu}(x) \sim  e^{ \frac{3}{2} x} , ~~~~~~~~x \to \infty,
\nonumber\\
&&    \overline{V}_{\mu}(x) \sim \frac{1}{\sqrt{x}}  , ~~~~~~~x \to 0,
\label{zmvas}
\end{eqnarray}
and 
\begin{eqnarray}
&& \overline{Z}_{\mu}(x) \sim  e^{ -\frac{3}{2} x}, ~~~~~~x \to \infty,
\nonumber\\
&&  \overline{Z}_{\mu}(x) \sim \frac{1}{\sqrt{x}}   , ~~~~~~~x \to 0.
\label{zmzas}
\end{eqnarray}

Consider now the normalization integrals. For the gauge field zero mode the 
normalization integral is  
\begin{equation}
\int_{0}^{\infty}~dx | \overline{{\cal A}}_{\mu}(x)|^2= |k_{1,\mu}|^2 \int_{0}^{\infty} L(x) P^2(x) dx.
\label{x1}
\end{equation}
From Eqs. (\ref{zmgas}) and from the explicit solutions where the asymptotics 
are realized, the integral of (\ref{x1}) always give a finite result. More specifically 
the integrand goes always as $x$ for $x\to 0$ and it is exponentially
suppressed for $x \to \infty$. It should be appreciated that 
this result has been derived only using the equations of motion of the background 
and of the fluctuations. In other words the asymptotic beahaviour  of the 
vortex solution is not specific of a given set of parameters but it is 
generic for the class of solutions discussed in Section III. 

For the gauge-invariant vector fluctuations of the metric the 
normalization integrals are:
\begin{eqnarray} 
&& \int_{0}^{\infty}~dx | \overline{V}_{\mu}(x)|^2 =  |c_{1,\mu}|^2 \int_{0}^{\infty} \frac{ d x}{M^4(x) L(x)},
\label{x2}\\
&&\int_{0}^{\infty}~dx | \overline{Z}_{\mu}(x)|^2 =\int_{0}^{\infty} 
\biggl[   | k_{1,\mu}|^2 L^3(x) +   |c_{2,\mu}|^2 \frac{M^4(x)}{L(x)} + 
2 k_{1,\mu} c_{1,\mu} M^2(x) L(x) \biggr] dx.
\label{x3}
\end{eqnarray}
Consider first Eq. (\ref{x2}).
Since for $x \to \infty$ the warp 
factors are exponentially decreasing the integrand of Eq. (\ref{x2}) 
diverges. This can be appreciated also from Eq. (\ref{zmvas}) whose 
square is the integrand appearing in (\ref{x2}).
Hence, the zero mode of $V_{\mu}$ is never  localized. 
Finally the second term of the integrand of Eq. (\ref{x3}) 
diverges as $1/x$ for $x\to 0$ leading to an integral which is 
logarithmically divergent in the same limit. Again, this 
can be also appreciated from Eqs. (\ref{zmzas}).
As a consequence, none of the graviphoton fields are localized
since their related normalization integrals are always divergent either 
close to the core of the defect, or at infinity. 

\renewcommand{\theequation}{5.\arabic{equation}}
\setcounter{equation}{0}
\section{Conclusions}

In this paper the vector and tensor fluctuations of the  six-dimensional 
Abelian-Higgs model have been considered. Thanks to the presence 
of a negative cosmological constant in the bulk the vortex solutions 
appearing in this framework lead to gravity localization and to a finite 
four-dimensional Planck mass. 
Since the four-dimensional Planck mass is finite, also the graviton zero mode is 
always localized. 

A different situation occurs for the vector fluctuations of the geometry whose 
normalization integrals lead to the following two conditions, namely, 
\begin{eqnarray}
&& |c_{1,\mu}|^2 \int_{0}^{\infty} \frac{ d x}{M^4(x) L(x)},
\label{x2c}\\
&& \int_{0}^{\infty} \biggl[   | k_{1,\mu}|^2 L^3(x) +   |c_{2,\mu}|^2 \frac{M^4(x)}{L(x)} + 
2 k_{1,\mu} c_{1,\mu} M^2(x) L(x) \biggr] dx.
\label{x3c}
\end{eqnarray}
Since the convergence of the four-dimensional Planck mass implies that 
\begin{equation}
\int_{0}^{\infty} M^2(x) L(x) dx, 
\end{equation}
is always finite, then the integral of (\ref{x2c}) will diverge at infinity.
Since the regularity of the geometry close to the core of the vortex implies that 
\begin{equation}
{\cal L}(x) = \sqrt{\lambda} v L(x) \sim x, ~~~~~M(x) \sim 1,
\end{equation}
for $x \to 0$, then the integral of (\ref{x3c}) will be divergent for $x\to 0$. 

An intriguing result, which should be further scrutinized,  holds for the gauge zero mode whose normalization integral 
implies that 
\begin{equation}
|k_{1,\mu}|^2 \int_{0}^{\infty} L(x) P^2(x) dx,
\end{equation}
should be finite. The local nature of the string-like defect 
demands, for the solutions localizing gravity presented in this paper, that 
$P(x) \to 1$ for $x\to 0$ and $P(x) \sim e^{- \sqrt{\alpha } x} $ for 
$x \to \infty$. This observation together with the regularity 
of the geometry in the core of the vortex, implies that the same solutions 
leading to gravity localization, also lead to the 
localization of the gauge zero mode. Notice that if the cosmological constant 
does not vanish, $\alpha = e^2/\lambda >1$. 
In fact, in the limit of zero cosmological constant (i.e. $\mu \to 0$ in our notations), 
$\alpha$ goes to $ 2$ \cite{mhm1} 
and the Bogomolnyi limit is recovered \footnote{The Bogomolnyi limit 
occurs, in our notations, for $\alpha =2$. Since $\alpha = 2 m^2_{V}/m_{H}^2$, 
$\alpha=2$ is the case when the vector boson and Higgs masses are equal.}.

It should be appreciated that the obtained results have been derived in general 
terms. First of all they are independent on the specific coordinate system 
since a fully gauge-invariant derivation as been employed. Second, the 
obtained results hold for all the class of backgrounds localizing gravity
in the six-dimensional Abelian Higgs model. In fact, even if specific background 
solutions
have been presented and used in order to illustrate the results, the zero 
modes have been computed without assuming any specific solution.

In order to interpret the localized gauge zero mode as an electromagnetic 
degree of freedom, the inclusion of fermions in the model is mandatory. 
This is the reason why we cannot claim that our findings support 
a mechanism for the localization 
of electromagnetic interactions on a string-like defect in higher dimensions. In
order to address precisely this point it would be interesting to 
use recent results concerning  fermionic degrees of freedom
on six-dimensional vortices \cite{tro1,tro2,lib}. 

\section*{Acknowledgments}
The author wishes to thank M. Shaposhnikov for inspiring discussions 
and insightful comments.

\newpage
\begin{appendix}
\renewcommand{\theequation}{A.\arabic{equation}}
\setcounter{equation}{0}
\section{Gauge-invariant fluctuations of the gravitating vortex}
In the following the explicit expressions 
of the various fluctuations needed for the derivations presented in the 
bulk of the paper will be reported. 
The background values of the Christoffel connections are 
\begin{eqnarray}
&& \overline{\Gamma}_{\alpha \rho}^{\beta} = \sqrt{\lambda} ~v~  H \delta_{\alpha}^{\beta},
\nonumber\\
&& \overline{\Gamma}_{\alpha\beta}^{\rho} = \sqrt{\lambda} ~ v~M^2 ~H \eta_{\alpha\beta}, 
\nonumber\\
&& \overline{\Gamma}_{\rho\theta}^{\theta} = \sqrt{\lambda}~v~ F ,
\nonumber\\
&& \overline{\Gamma}_{\theta\theta}^{\rho} = - \sqrt{\lambda}~v~ L^2 F .
\label{backch}
\end{eqnarray}
Notice that $\rho$ and $\theta$ denote the two transverse coordinates whereas the 
other Greek letters label the four space-time coordinates. Notice, furthermore, that 
as indicated in the bulk of the paper, the prime denotes the derivation with respect to 
$ x =\sqrt{\lambda} v \rho$, and the overdot the derivative with respect to $\theta$.  
As it is apparent from expressions of the 
evolution equations of the fluctuations, the use of the $x$ and $\theta$ 
(instead of $\rho$ and $\theta$) 
 makes the whole discussion simpler.

Using Eq. (\ref{lorf2}) the first order fluctuations of the 
Christoffel connections are easily obtained \footnote{ While in the main text the first order 
fluctuation of a given tensor with respect to (\ref{lorf2}) has been indicated 
(for sake of simplicity) by $\delta$, in this Appendix the notation $\delta^{(1)}$ will 
be followed. Notice moreover, that the factors $\sqrt{\lambda} v$ appear 
because the derivatives (denoted with the prime) are taken with respect to the rescaled 
bulk radius $x$.} :
\begin{eqnarray}
&& \delta^{(1)}\Gamma_{\mu \nu}^{\rho} = \sqrt{\lambda} ~v~ M^2 [ H_{\mu\nu}'  + 2 H H_{\mu\nu} ] - 
M \partial_{(\mu}D_{\nu)},
\nonumber\\
&& \delta^{(1)} \Gamma_{\rho \mu}^{\rho} = \sqrt{\lambda}~v~ M~H D_{\mu},
\nonumber\\
&& \delta^{(1)} \Gamma_{\mu\theta}^{\rho} = \frac{\sqrt{\lambda} ~v}{2} ~M L [ Q_{\mu}' + ( H + F) Q_{\mu} ] - 
\frac{M}{2} \dot{D}_{\mu} ,
\nonumber\\
&& \delta^{(1)} \Gamma_{\mu\nu}^{\theta}  = \frac{M^2}{L^2} \dot{H}_{\mu\nu} - \frac{M}{L} \partial_{(\mu} Q_{\nu)},
\nonumber\\
&& \delta^{(1)} \Gamma_{\mu\rho}^{\theta} = \frac{ M}{2 L^2} \dot{D}_{\mu}  - \sqrt{\lambda} v 
\frac{M}{2 L} [ Q_{\mu}' + Q_{\mu} ( F- H)]
\nonumber\\
&&  \delta^{(1)} \Gamma_{\alpha\beta}^{\mu} = - \sqrt{\lambda}~ v ~ H M D^{\mu} \eta_{\alpha\beta} 
+ ( \partial_{\alpha} H_{\beta}^{~\mu} + \partial_{\beta} H_{\alpha}^{~\mu}  - \partial^{\mu} H_{\alpha\beta}),
\nonumber\\
&& \delta^{(1)} \Gamma_{\alpha \rho}^{\mu} = \frac{1}{2 M} ( \partial_{\alpha} D^{\mu} - \partial^{\mu} D_{\alpha} )
+ \sqrt{\lambda} ~ v~{H_{\alpha}^{\mu}}',
\nonumber\\
&& \delta^{(1)} \Gamma_{\theta\theta}^{\mu} = \frac{L}{M} \dot{Q}^{\mu} + \sqrt{\lambda ~ v}\frac{L^2}{M} ~ F D^{\mu}, 
\nonumber\\
&& \delta^{(1)} \Gamma_{\alpha\theta}^{\mu} = \frac{L}{2 M} ( \partial_{\alpha} Q^{\mu} - \partial^{\mu} Q_{\alpha} )
+ \dot{H}_{\alpha}^{~\mu}, 
\nonumber\\
&& \delta^{(1)} \Gamma_{\rho\rho}^{\mu} = \frac{\sqrt{\lambda} ~v}{M}[ {D^{\mu}}' + H D^{\mu}],
\nonumber\\
&& \delta^{(1)} \Gamma_{\theta\rho}^{\mu} = \sqrt{\lambda}~ v~ 
\frac{L}{2 M} [ {Q^{\mu}}' + ( H- F) Q^{\mu}] + 
\frac{\dot{D}^{\mu}}{2 M},
\label{chfl}
\end{eqnarray}
where, for short, 
\begin{equation}
H_{\mu\nu} = h_{\mu\nu} + \partial_{(\mu}f_{\nu)}.
\label{HMN}
\end{equation}
With the use of  Eqs. (\ref{backch}) and 
 (\ref{chfl}), Eq. (\ref{ricci}) allows the explicit determination 
 of the first order Ricci fluctuations:
\begin{eqnarray}
 \delta^{(1)} R_{\mu\nu} &=& \lambda ~ v^2 M^2 [ H_{\mu\nu}'' + ( 4 H + F) H_{\mu\nu}' + 
H_{\mu\nu} ( 2 H ' + 8 H^2 + 2 H F) ] + \frac{M^2}{L^2} \ddot{H}_{\mu\nu}
\nonumber\\
&-& \partial_{\alpha}\partial^{\alpha} H_{\mu\nu} + ( \partial_{\alpha} \partial_{\mu} H_{\nu}^{\alpha} + 
\partial_{\alpha} \partial_{\nu} H_{\mu}^{\alpha} + \partial_{\nu}\partial^{\alpha} H_{\alpha\mu} -
\partial_{\nu}\partial_{\alpha} H_{\mu}^{\alpha}),
\nonumber\\
&-& \sqrt{ \lambda}~v ~ M [ \partial_{(\mu} D_{\nu)}' + ( 3 H + F) \partial_{(\mu} D_{\nu)}]
- \frac{M}{L} \partial_{(\mu }\dot{Q}_{\nu)},
\label{mn}\\
 \delta^{(1)} R_{\mu\rho} &=& \frac{M}{ 2 L^2} \ddot{D}_{\mu} - \frac{M}{2 L}\sqrt{\lambda}~v~ 
[ \dot{Q}'_{\mu} + ( F - H)  \dot{Q}_{\mu}] - \frac{1}{2 M} \partial_{\alpha}\partial^{\alpha} D_{\mu} 
\nonumber\\
&+& 
\lambda ~v^2 M [ H' + 4 H^2 + H F] D_{\mu}   + \sqrt{\lambda}~ v H_{\mu\alpha}', 
\label{mr}\\
 \delta^{(1)} R_{\mu\theta} &=& \partial^{\alpha} \dot{H}_{\alpha\mu} 
- \frac{L}{2 M} \partial_{\alpha} \partial^{\alpha} Q_{\mu} - \frac{ \sqrt{\lambda} ~v}{2} ~ M [ \dot{D}_{\mu}' + ( 5 H - F) 
\dot{D}_{\mu} ] 
\nonumber\\
&+& \lambda ~v^2~\frac{M L}{2} [ Q_{\mu}'' + ( 4 H + F) Q_{\mu}' + 
( H' + F' + 3 H^2 + 7 HF)Q_{\mu}]. 
\label{mth}
\end{eqnarray}
The first order fluctuations of $\tau_{A B}$ defined in Eq. (\ref{cont3}) 
and appearing at the right hand side of Eqs. (\ref{I})--(\ref{III}) can be also 
computed and they are
\begin{eqnarray}
\kappa \delta^{(1)} \tau_{\mu\nu} &=&  \lambda ~v^2 M^2 \biggl[ - \mu + \frac{\nu}{2 \alpha} \frac{P'^2}{{\cal L}^2} - 
\frac{\nu}{4} (f^2 -1)^2 \biggr] H_{\mu\nu},
\label{taupmn}\\ 
\kappa \delta^{(1)} \tau_{\mu\rho} &=& \lambda~v^2 M \biggl[ - \frac{\mu}{2} + \frac{\nu}{4 \alpha} \frac{P'^2}{{\cal L}^2} - 
\frac{\nu}{8} (f^2 -1)^2 \biggr] D_{\mu} + \sqrt{\lambda} ~v~
 \frac{\nu}{\alpha} \frac{P'}{{\cal L}^2}  \dot{{\cal A}}_{\mu},
\label{taupmr}\\
\kappa \delta^{(1)} \tau_{\mu\theta} &=& \lambda v^2 M L  
\biggl[ - \frac{\mu}{2} + \frac{\nu}{4 \alpha} \frac{P'^2}{{\cal L}^2} - 
\frac{\nu}{8} (f^2 -1)^2 \biggr] Q_{\mu} 
- \frac{\nu}{\alpha} P' {\cal A}_{\mu}' - \nu P f^2 {\cal A}_{\mu}.
\label{taupmth}
\end{eqnarray}
In order to obtain the evolution equations written explicitly in terms of the 
gauge-invariant fluctuations we recall that while $h_{\mu\nu}$ and ${\cal A}_{\mu}$ 
are already gauge-invariant, $f_{\mu}$, $D_{\mu}$ and $ Q_{\mu}$ change for infinitesimal coordinate 
transformation according to Eqs. (\ref{fl})--(\ref{zeta2}).

The evolution equations of the fluctuations will now be written in fully gauge-invariant terms. 
The strategy will be the, in short, the following. By using Eqs. (\ref{gia}) and (\ref{gib}), 
the first order fluctuations of the Ricci tensors can be expressed in terms
of a gauge-invariant part plus a gauge-dependent piece. The same procedure 
can be carried on in the case of the fluctuations of the energy-momentum tensor.
When the Einstein's equations are explicitly written to first 
order in the amplitude of the metric fluctuations as, 
\begin{equation}
\delta^{(1)} R_{A B} = \kappa \delta^{(1)}\tau_{ A B}
\label{pertfirst}
\end{equation}
the gauge-dependent parts vanish, identically, by using 
the equations of motion of the background reported in Eqs. (\ref{f1})--(\ref{l1}).

From Eqs. (\ref{zeta1})--(\ref{zeta2}) we can write 
\begin{eqnarray}
D_{\mu} &=& V_{\mu} + \sqrt{\lambda} ~v ~M f_{\mu}'
\nonumber\\
Q_{\mu} &=& Z_{\mu} + \frac{M}{L} \dot{f}_{\mu}.
\label{QD}
\end{eqnarray}
Using Eqs. (\ref{QD}) into Eqs. (\ref{mn})--(\ref{mth}) and (\ref{taupmn})--(\ref{taupmth})
the following first order fluctuations can be obtained
\begin{eqnarray}
\delta^{(1)} R_{\mu\nu} &=& \lambda v^2 M^2 [ h_{\mu\nu}'' + ( 4 H + F) h_{\mu\nu}' + 
( 2 H' + 8 H^2 + 2 H F) h_{\mu\nu} ] + \frac{M^2}{L^2} \ddot{h}_{\mu\nu} 
- \partial_{\alpha}\partial^{\alpha} h_{\mu\nu} 
\nonumber\\
&-& \sqrt{\lambda}~v~ M [ \partial_{(\mu}V_{\nu)}' + ( 3 H + F) \partial_{(\mu}V_{\nu)}] 
- \frac{M}{L} \partial_{(\mu} \dot{Z}_{\nu)} 
\nonumber\\
&+& \lambda v^2 M^2 ( 2 H' + 8 H^2 + 2 H F) 
 \partial_{(\mu}f_{\nu)},
\label{gimn}\\
\delta^{(1)} R_{\mu\rho} &=& \frac{M}{2 L^2} \ddot{V}_{\mu} 
- \frac{M}{2 L} \sqrt{\lambda} ~v [ \dot{Z}_{\mu}' + \dot{Z}_{\mu} ( F - H)] 
\nonumber\\
&-& 
\frac{1}{2 M} \partial_{\alpha} \partial^{\alpha} V_{\mu} + 
\lambda v^2 M V_{\mu} [ H' + 4 H^2 + H F] 
\nonumber\\
&+& (\sqrt{\lambda} v)^3 ~M^2 [ H' + 4 H^2 + H F] f_{\mu}' ,
\label{gimr}\\
\delta^{(1)} R_{\mu\theta} &=& - \frac{L}{2 M} \partial_{\alpha} \partial^{\alpha} Z_{\mu}+ 
\lambda v^2 \frac{M L}{2}[ Z_{\mu}'' + ( 4 H + F) Z_{\mu}' + 
Z_{\mu} ( H' + F' + 3 H^2 + 7 H F) ]
\nonumber\\
&-& 
\frac{\sqrt{\lambda} v}{2} M[ \dot{V}_{\mu}' + ( 5 H - F) \dot{V}_{\mu}]
+\lambda v^2 M^2 (  H' + 4 H^2 +  H F) \dot{f}_{\mu} ,
\label{gimth}
\end{eqnarray}
and 
\begin{eqnarray}
\kappa \delta^{(1)} \tau_{\mu\nu} &=& \lambda v^2 M^2 ( 2 H' + 8 H^2 + 2 H F) h_{\mu\nu} 
+ \lambda v^2 M^2 \partial_{(\mu}f_{\nu)} ( 2 H' + 8 H^2 + 2 H F) ,
\label{gitau1}\\
\kappa \delta^{(1)} \tau_{\mu\rho} &=& \lambda v^2 M (  H' + 4 H^2 +  H F) V_{\mu} +
\frac{\nu}{\alpha} \sqrt{\lambda} ~v \frac{P'}{{\cal L}^2} \dot{{\cal A}}_{\mu} 
\nonumber\\
&+& (\sqrt{\lambda} v)^3 M^2 (  H' + 4 H^2 + 
 H F) f_{\mu}' ,
\label{gitau2}\\
\kappa \delta^{(1)} \tau_{\mu\theta} &=& \lambda v^2 M L ( H' + 4 H^2 + H F) Z_{\mu}
- \frac{\nu}{\alpha} P' {\cal A}_{\mu}' - \nu P f^2 {\cal A}_{\mu} 
\nonumber\\
&+& 
\lambda v^2 M^2 [ H' + 4 H^2 + H F] \dot{f}_{\mu}.
\label{gitau3}
\end{eqnarray}
Notice that in order to write  Eqs. (\ref{gimn})--(\ref{gimth}) and (\ref{gitau1})--(\ref{gitau3})
the following expression (following from the sum of Eqs. (\ref{m2}) and (\ref{l1}) ) 
has been used:
\begin{equation}
2 H' + 8 H^2 + 2 H F = - \mu - \frac{\nu}{4} (f^2 - 1)^2 + \frac{\nu}{2 \alpha} 
\frac{P'^2}{{\cal L}^2}.
\end{equation}
In each of  Eqs. (\ref{gimn})--(\ref{gimth}) and (\ref{gitau1})--(\ref{gitau3}) the last 
term is not gauge-invariant. However, imposing Eq. (\ref{pertfirst}) all the 
gauge-dependent pieces cancel and, at the end, the following system of equations is obtained:
\begin{eqnarray}
&& V_{\mu}' + ( 3 H + F) V_{\mu}  + \frac{\dot{Z}_{\mu}}{{\cal L}} =0,
\label{v1da}\\
&& \frac{\ddot{V}_{\mu}}{{\cal L}^2} - \frac{2}{m_{H}^2 ~M^2}\Box V_{\mu} - 
\frac{\dot{Z}'_{\mu}}{{\cal L}} + ( H - F) \frac{\dot{Z}_{\mu}}{{\cal L}} = 
2 \frac{L}{M} \frac{\nu}{\alpha} \frac{P'}{{\cal L}^2} 
\frac{\dot{{\cal A}}_{\mu}}{{\cal L}},
\label{v2da}\\
&& Z_{\mu}'' + ( 4 H + F) Z_{\mu}' + ( F' - H' + 5 H F - 5 H^2) Z_{\mu} 
- \frac{ 2}{ m_{H}^2 ~M^2} \Box Z_{\mu}
\nonumber\\
&& - \frac{1}{{\cal L}} \biggl[ \dot{V}_{\mu}' + ( 5 H 
- F) \dot{V}_{\mu}\biggr] + 2 \frac{L}{M} \biggl( \frac{\nu}{\alpha} 
\frac{P'}{{\cal L}^2} {\cal A}_{\mu}' + \frac{\nu P f^2}{{\cal L}^2} 
{\cal A}_{\mu} \biggr) =0,
\label{v3da}
\end{eqnarray}
for the vector fluctuations and
\begin{equation}
h_{\mu\nu}'' + \frac{\ddot{h}_{\mu\nu}}{{\cal L}^2} + ( 4 H + F) h_{\mu\nu}' 
- \frac{2}{m_{H}^2 M^2} \Box h_{\mu\nu} =0,
\label{sp2}
\end{equation}
for the tensor fluctuations. 

This system of equations has to be supplemented with the fluctuation of the 
gauge field equation reported in Eq. (\ref{Aper}). The pure vector component 
of the perturbed evolution equation for the gauge field reads: 
\begin{eqnarray}
&& {\cal A}_{\mu}'' 
+ \frac{\ddot{{\cal A}}_{\mu}}{{\cal L}^2} - \frac{2}{m_{H}^2~M^2} \Box
{\cal A}_{\mu} + ( 2 H + F) {\cal A}_{\mu}'
\nonumber\\
&& - P' \frac{M}{L} \biggl[ Q_{\mu}' - (H- F) Q_{\mu} - \frac{\dot{D}_{\mu}}{{\cal L}}
\biggr] - \alpha {\cal A}_{\mu} f^2 =0.
\label{pgf}
\end{eqnarray}
In this equation the terms containing ${\cal A}_{\mu}$ are gauge-invariant. The terms 
containing $Q_{\mu}$ and $D_{\mu}$ are not automatically gauge-invariant. 
However, using Eqs. (\ref{QD}) the following result holds :
\begin{equation} 
Q_{\mu}' - (H- F) Q_{\mu} - \frac{\dot{D}_{\mu}}{{\cal L}} =  
Z_{\mu}' - (H- F) Z_{\mu} - \frac{\dot{V}_{\mu}}{{\cal L}},
\end{equation}
and the dependence upon $f_{\mu}$ disappears, as it should. Therefore the final equation for 
the perturbed gauge field fluctuation is simply 
\begin{eqnarray}
&& {\cal A}_{\mu}'' 
+ \frac{\ddot{{\cal A}}_{\mu}}{{\cal L}^2} - \frac{2}{m_{H}^2~M^2} \Box
{\cal A}_{\mu} + ( 2 H + F) {\cal A}_{\mu}'
\nonumber\\
&& - P' \frac{M}{L} \biggl[ Z_{\mu}' - (H- F) Z_{\mu} - \frac{\dot{V}_{\mu}}{{\cal L}}
\biggr] - \alpha {\cal A}_{\mu} f^2 =0.
\label{pgf2}
\end{eqnarray}
\end{appendix}

\end{document}